# Atomic Security Concerns: Towards a Foundational Language for Security Knowledge Representation

**Hanane Houmani**

*Secure-It Labs*
*Montreal, QC, Canada*
hanane.houmani@secureitlab.ca

## *Abstract*

*Cybersecurity governance, risk management, compliance, and security architecture rely on numerous concepts, including security properties, domains, controls, risks, implementation measures, capabilities, and compliance requirements. Although these concepts are widely used, they are often represented through heterogeneous abstractions and vocabularies, making relationships among them difficult to express consistently and leaving much of the associated reasoning dependent on expert knowledge.*

*This paper introduces the concepts of Security Concerns and Atomic Security Concerns (ASCs) and explores the hypothesis that cybersecurity knowledge may be represented through a finite set of foundational concerns. Inspired by the role of atoms in chemistry, the proposed approach views cybersecurity concepts as compositions of more fundamental concerns rather than as independent abstractions. Hence, Atomic Security Concerns could be viewed as a common abstraction layer for the representation of cybersecurity concepts, independently of their domain, nature, or level of granularity.*

*Building upon this perspective, the paper explores how Atomic Security Concerns may provide the vocabulary of a cybersecurity language and how cybersecurity concepts may be represented as concern sets supporting explicit relationships among concepts. The contribution of this work is intentionally conceptual: it introduces a framework and research hypothesis for investigating the foundations of a common language for cybersecurity knowledge representation, reasoning, and future formalization.*

## 1 Introduction

Cybersecurity governance, risk management, compliance, and security architecture rely on a diverse set of cybersecurity concepts, including security properties, security domains, security controls, security requirements, security risks, and security architectures. These concepts are extensively employed across standards, regulatory frameworks, risk assessment methodologies, and architectural practices. Despite their widespread adoption, they are generally defined through distinct vocabularies, conceptual structures, and levels of abstraction, making it difficult to establish explicit, consistent, and traceable relationships among them.

In practice, cybersecurity professionals routinely reason about relationships between security risks and security controls, security controls and security domains, security requirements and technical implementations, or organizational capabilities and security objectives. However, such relationships are rarely represented through a unified conceptual framework. Instead, they often depend on tacit knowledge acquired through experience, organizational practices, and familiarity with multiple standards and frameworks. Consequently, cybersecurity reasoning remains difficult to formalize, communicate, automate, and systematically reproduce across organizational and technological contexts.

This observation raises a fundamental research question: **does cybersecurity possess a foundational language capable of representing its knowledge structures through a common and coherent conceptual model?**

Many mature scientific disciplines rely on a finite set of foundational elements from which increasingly complex structures can be constructed. Natural languages are built upon alphabets that enable the construction of words, sentences, and documents. Similarly, chemistry relies on atoms as foundational elements from which molecules and more complex forms of matter are derived. In both cases, a relatively small set of foundational elements provides the expressive capacity required to represent a substantially larger body of knowledge.

This paper investigates whether a similar principle may exist within cybersecurity. More specifically, it explores the hypothesis that cybersecurity knowledge can be represented through a limited set of **Atomic Security Concerns (ASCs)** that serve as candidate foundational elements of a cybersecurity language. Under this hypothesis, cybersecurity concepts such as security properties, security domains, security controls, security requirements, security risks, and security architectures are viewed not as independent abstractions, but as compositions of more fundamental security concerns.

The objective of this work is not to establish an exhaustive catalog of Atomic Security Concerns. Rather, it seeks to introduce a conceptual framework for investigating the existence of a cybersecurity language founded upon a finite set of foundational elements. More specifically, this work seeks to characterize the properties that such foundational elements should possess. It is proposed that Atomic Security Concerns satisfy two essential properties. First, they should exhibit **irreducibility**, meaning that they cannot be decomposed into more fundamental security concerns without loss of security semantics. Second, they should exhibit **expressiveness**, meaning that a sufficiently complete set of concerns should provide the representational capacity required to describe the principal cybersecurity concepts encountered within cybersecurity governance, risk management, compliance, and security architecture.

Based on these principles, this paper makes five primary contributions:

1. It introduces the hypothesis that Atomic Security Concerns may serve as foundational elements of a cybersecurity language.
2. It proposes irreducibility and expressiveness as the defining properties of Atomic Security Concerns.
3. It presents an initial illustrative taxonomy of candidate Atomic Security Concerns.
4. It demonstrates how common cybersecurity concepts can be represented as compositions of Atomic Security Concerns.
5. It discusses the implications of this perspective for cybersecurity knowledge representation, traceability, reasoning, and future automation.

The remainder of this paper is organized as follows. Section 2 reviews related work and discusses existing approaches to cybersecurity knowledge representation. Section 3 introduces the concept of Atomic Security Concerns and presents their definition, atomicity principles, and identification criteria. Section 4 presents an initial taxonomy of Atomic Security Concerns. Section 5 introduces the proposed composition model and demonstrates how Atomic Security Concerns can be used to represent security properties, security capabilities, security functions, security domains, and security controls. Section 6 provides illustrative examples and validation scenarios. Finally, Sections 7 through 9 discuss implications and limitations of the proposed approach, identify future research directions, and present the conclusions of this work.

## 2 Related Work

Cybersecurity knowledge has traditionally been structured through multiple complementary perspectives, including security properties, security controls, cybersecurity ontologies, and framework mapping initiatives. These approaches provide mechanisms for describing, implementing, organizing, and relating cybersecurity concepts. However, they generally do not address whether such concepts can be reduced to a common set of foundational elements from which broader cybersecurity knowledge structures may be derived.

### 2.1 Security Properties

Security properties constitute one of the most established approaches to describing cybersecurity objectives. The Confidentiality, Integrity, and Availability (CIA) model remains a widely adopted conceptual framework for expressing fundamental security objectives. Additional properties, including authenticity, accountability, non-repudiation, privacy, and resilience, have subsequently been introduced to address broader cybersecurity concerns.

Although security properties provide a high-level description of desired security outcomes, they do not provide a mechanism for representing the relationships among security properties, security controls, security domains, security requirements, security risks, and security architectures. Consequently, they are generally treated as objectives or characteristics of secure systems rather than as foundational elements of a cybersecurity representation model.

## 2.2 Control-Based Frameworks

Security standards and frameworks frequently organize cybersecurity knowledge through collections of security controls. Frameworks such as NIST Special Publication 800-53 [2], ISO/IEC 27001 [1,3], and CIS Controls classify controls into categories intended to address specific security objectives and risks.

This approach provides practical guidance for governance, risk management, compliance, and security implementation activities. However, control frameworks primarily define security measures rather than the conceptual structures from which such measures originate. Multiple controls may contribute to the same security objective, while a single control may support several objectives simultaneously. As a result, control frameworks provide limited support for explaining the underlying conceptual relationships among security controls, security risks, security requirements, security properties, and security architectures.

## 2.3 Cybersecurity Ontologies

Cybersecurity ontologies have been proposed to improve the formal representation, integration, and interoperability of cybersecurity knowledge. Existing work has modeled concepts such as assets, threats, vulnerabilities, controls, incidents, and compliance requirements, together with the relationships connecting these concepts. Examples include security ontology surveys by Souag et al. [5] and ontology engineering efforts such as the cybersecurity ontology proposed by Oltramari et al. [6], which seek to provide formal semantic structures for cybersecurity knowledge.

More recently, ontology-based approaches have been extended to support the integration of heterogeneous cybersecurity knowledge sources. Examples include the Unified Cybersecurity Ontology (UCO) and more recent work integrating MITRE ATT&CK [9], D3FEND [10], CWE, and CVE [4] within common semantic frameworks. These approaches aim to improve interoperability, semantic consistency, automated reasoning, and knowledge sharing across cybersecurity repositories.

Ontology-based approaches [5,6,7,8] focus on defining concepts and formalizing relationships among those concepts. Their primary objective is therefore the representation and integration of cybersecurity knowledge within coherent semantic structures. However, cybersecurity ontologies generally assume the existence of the cybersecurity concepts being represented. They describe how concepts relate to one another but do not typically investigate whether those concepts can themselves be decomposed into more fundamental semantic elements.

The research question addressed in this work differs from that of ontology engineering. Rather than focusing on the representation of cybersecurity concepts and their relationships, it investigates whether a finite set of irreducible security concerns may exist beneath those concepts. Under this perspective, concepts commonly represented in cybersecurity ontologies may themselves be expressible as compositions of more fundamental concerns.

Consequently, Atomic Security Concerns should not be viewed as an alternative to cybersecurity ontologies but rather as a potentially complementary abstraction layer. If such concerns can be identified and validated, they may provide foundational semantic elements upon which future cybersecurity ontologies can be constructed.

## 2.4 Cross-Framework Mapping and Interoperability

A complementary line of research focuses on establishing correspondences among cybersecurity standards, regulations, frameworks, and methodologies. Crosswalks, mappings, and harmonization initiatives seek to improve consistency, interoperability, and traceability by establishing relationships among controls, requirements, and concepts defined by different frameworks.

These approaches recognize that many cybersecurity frameworks address similar concerns through different organizational structures and vocabularies. While framework mappings facilitate alignment and interoperability, they remain dependent on the concepts defined within the frameworks being mapped. Consequently, they do not generally address the existence of more fundamental semantic elements that may underlie those concepts.

## 2.5 Research Gap and Positioning of This Work

The reviewed approaches demonstrate that cybersecurity knowledge can be organized according to multiple perspectives, including security properties, security controls, ontology models, and framework mappings. These approaches contribute to the description, implementation, representation, integration, and interoperability of cybersecurity knowledge.

However, they generally address cybersecurity concepts as predefined entities. Security properties define objectives, control frameworks prescribe measures, ontologies formalize concepts and relationships, and mapping initiatives establish correspondences among existing frameworks. Comparatively little attention has been devoted to identifying whether such concepts originate from a smaller set of irreducible security concerns.

More specifically, existing ontology-based approaches focus on defining concepts, establishing semantic relationships, and enabling interoperability among heterogeneous knowledge sources. Similarly, taxonomy and framework mapping initiatives seek to classify, organize, or align existing cybersecurity concepts. In contrast, relatively little research has investigated whether cybersecurity concepts themselves can be decomposed into a finite set of more fundamental concerns.

This observation motivates the following research question:

*"Can cybersecurity concepts be represented as compositions of a finite set of irreducible and expressive security concerns?"*

To address this question, this work investigates the hypothesis that cybersecurity concepts may be decomposed into Atomic Security Concerns (ASCs). Unlike security properties, which define objectives, control frameworks, which prescribe measures, ontology models, which define and relate concepts, or framework mappings, which align existing conceptual structures, the proposed approach focuses on identifying candidate foundational elements from which cybersecurity concepts may be constructed.

The objective is not to replace existing cybersecurity standards, frameworks, taxonomies, or ontologies. Instead, it is to explore whether a finite set of irreducible and expressive security concerns can provide a common foundation for representing higher-level cybersecurity concepts. Under this hypothesis, concepts such as security properties, security capabilities, security functions, security domains, and security controls may be represented as compositions of Atomic Security Concerns, supporting the construction of a formal language capable of expressing cybersecurity knowledge in a consistent and traceable manner.

## 3 Atomic Security Concerns

The hypothesis proposed in this paper is that cybersecurity knowledge may be represented through a finite set of foundational elements referred to as **Atomic Security Concerns (ASCs)**. Hence, Atomic Security Concerns (ASCs) are proposed as a common abstraction layer for the representation of cybersecurity concepts, independently of their domain, nature, or level of granularity.

Before defining Atomic Security Concerns, it is first necessary to establish the broader notion of a **Security Concern**, as Atomic Security Concerns represent a specific category of Security Concerns characterized by the property of atomicity.

## 3.1 Security Concerns

Cybersecurity knowledge is commonly expressed through concepts such as security properties, security controls, security requirements, security risks, security capabilities, security functions, and security architectures. Although these concepts differ in purpose, scope, and level of abstraction, they frequently address underlying security-related matters that motivate cybersecurity decisions, objectives, and activities.

In this work, a **Security Concern** is defined as a security-related matter, objective, condition, capability, property, or expectation that stakeholders seek to establish, preserve, verify, enforce, or protect within a cybersecurity context.

Security Concerns are independent of specific technologies, implementation mechanisms, standards, frameworks, or organizational structures. Instead, they represent the underlying security considerations that recur across different cybersecurity concepts and contexts.

For example, Authentication may be analyzed as a composite Security Concern. In order to authenticate an entity, a system must first establish the Identity being claimed and then verify the Authenticity of that claim. Under this interpretation, Authentication can be represented as the combination of the concerns Identity and Authenticity.

Similarly, Confidentiality may be viewed as a composite concern. Confidentiality requires information to remain secret from unauthorized parties while ensuring that access to the information is restricted to authorized entities. Under this interpretation, Confidentiality can be represented as the combination of the concerns Secret and Access.

Non-Repudiation provides a third example. Demonstrating that an action cannot later be denied generally requires the ability to associate the action with an Identity, to maintain Traceability of the action, and to preserve Evidence supporting that association. Under this interpretation, non-repudiation can be represented as the combination of Identity, Traceability, and Evidence.

These examples suggest that some Security Concerns may themselves be composed of other concerns. Consequently, not every Security Concern should be assumed to be fundamental. This observation motivates the introduction of Atomic Security Concerns.

Table 1 provides illustrative examples of Security Concerns and potential decompositions into more fundamental concerns.

| Composite Security Concern | Illustrative Constituent Concerns | Rationale |
|---|---|---|
| Authentication | Identity, Authenticity | Verification requires both an asserted identity and validation of that identity. |
| Confidentiality | Secret, Access | Information must remain secret and access must be restricted. |
| Non-Repudiation | Identity, Traceability, Evidence | Attribution requires identification, activity traceability, and supporting evidence. |
| Privileged Access Management | Identity, Privilege, Access | Privileged actions require identification, privilege assignment, and access enforcement. |
| Secure Communications | Secret, Integrity, Trust | Communications require protection of content, preservation of integrity, and trusted endpoints. |

**Table 1. Illustrative examples of Security Concern decomposition**

The purpose of these examples is purely illustrative. The validity of such decompositions and the atomic status of the underlying concerns are examined through the principles and criteria introduced later in this paper.

## 3.2 Atomic Security Concerns

An **Atomic Security Concern (ASC)** is a Security Concern that cannot be further decomposed into more fundamental Security Concerns without loss of security semantics.

The notion of atomicity adopted in this work is semantic rather than technical. A concern is not considered atomic because it is simple, widely used, or represented at a lower level of abstraction. Instead, a concern is considered atomic when no decomposition into more fundamental Security Concerns can preserve its cybersecurity meaning.

For illustrative purposes, consider the concept of **Authentication**. Authentication is a meaningful cybersecurity concept and therefore constitutes a Security Concern. However, if authentication can be systematically represented as the combination of **Identity** and **Authenticity**, then Authentication should be regarded as a composite Security Concern rather than an Atomic Security Concern.

Similarly, if **Confidentiality** can be represented as the combination of **Secret** and **Access**, then Confidentiality is better understood as a higher-level Security Concern derived from more fundamental concerns. Likewise, if **Non-Repudiation** can be represented through the concerns **Identity**, **Traceability**, and **Evidence**, it should also be regarded as a composite Security Concern.

By contrast, concerns such as **Identity**, **Authenticity**, **Secret**, **Traceability**, **Evidence**, **Privilege**, or **Trust** may be considered candidate Atomic Security Concerns if they cannot be decomposed into more fundamental Security Concerns without loss of security semantics. Whether these concerns are truly atomic remains an open question and requires evaluation through the identification criteria proposed in this work.

The proposed model assumes the following relationship:

1. Every Atomic Security Concern is a Security Concern.
2. A Security Concern may be either atomic or composite.
3. A composite Security Concern can be represented as a composition of one or more Atomic Security Concerns.

Under this model, Authentication, Confidentiality, and Non-Repudiation are treated as composite Security Concerns because they can be represented through combinations of more fundamental concerns. In contrast, Identity, Authenticity, Secret, Traceability, and Evidence are treated as candidate Atomic Security Concerns pending evaluation against the atomicity principles introduced in the following section.

The relationship can therefore be summarized as:

*Atomic Security Concerns ⊂ Security Concerns*

All Atomic Security Concerns are Security Concerns, but not all Security Concerns are Atomic Security Concerns.

The distinction between Security Concerns and Atomic Security Concerns is analogous to the distinction between molecules and atoms in chemistry. A molecule may be decomposed into more fundamental elements while preserving its scientific interpretation. Atoms, by contrast, represent the smallest units considered fundamental within the model. Similarly, composite Security Concerns may be represented as combinations of more fundamental concerns, while Atomic Security Concerns represent candidate foundational elements from which cybersecurity concepts may be derived.

Under the hypothesis proposed in this paper, Atomic Security Concerns constitute the basic vocabulary of a cybersecurity language. Higher-level concepts such as security properties, security capabilities, security functions, security domains, and security controls may therefore be represented as compositions of one or more Atomic Security Concerns. The following sections introduce the principles of atomicity and the criteria used to identify candidate Atomic Security Concerns.

## 3.3 Atomicity Principles

The identification of Atomic Security Concerns requires a definition of atomicity that can be applied consistently across cybersecurity concepts. Without such principles, the distinction between composite Security Concerns and Atomic Security Concerns would remain subjective and dependent on individual interpretation.

In this work, atomicity is evaluated from a semantic perspective rather than a technical, organizational, or implementation perspective. A Security Concern is not considered atomic because it is frequently used, because it appears at a low level of abstraction, or because it is associated with a specific technology or control. Instead, atomicity depends on whether the concern can be meaningfully represented through more fundamental Security Concerns without loss of cybersecurity semantics. Three principles are proposed to evaluate the atomicity of a Security Concern.

### 3.3.1 Principle 1: Decomposability

A Security Concern should not be considered atomic if it can be consistently represented as a composition of other Security Concerns.

**Example: Authentication**

Authentication may be represented as:

*Authentication = Identity + Authenticity*

Authentication therefore represents a composite Security Concern because its meaning can be expressed through more fundamental concerns.

**Example: Non-Repudiation**

Non-Repudiation may be represented as:

*Non-Repudiation = Identity + Traceability + Evidence*

The concern can therefore be decomposed into constituent concerns while preserving its cybersecurity meaning.

If a concern can be represented through a meaningful combination of other Security Concerns, it should be regarded as a composite Security Concern rather than an Atomic Security Concern.

### 3.3.2 Principle 2: Preservation of Security Semantics

A decomposition is only valid if the resulting concerns preserve the original cybersecurity meaning of the concept being decomposed.

**Example: Valid Decomposition**

Authentication may be decomposed into:

*Identity + Authenticity*

The resulting concerns collectively preserve the essential meaning of authentication because authentication requires both an asserted identity and verification that the identity claim is genuine.

**Example: Invalid Decomposition**

Authentication cannot reasonably be decomposed into:

*Identity + Access*

Although access control frequently follows authentication, the notion of access does not capture the semantic meaning of verifying the authenticity of an identity claim.

Similarly, Confidentiality cannot be reduced to:

*Secret*

because confidentiality also requires restrictions on access to information.

Hence, a decomposition is acceptable only when the resulting concerns preserve the complete cybersecurity meaning of the original concern. Decompositions that omit essential semantics should be rejected.

### 3.3.3 Principle 3: Reusability Across Concepts

A candidate Atomic Security Concern should appear repeatedly across multiple cybersecurity concepts and contexts.

**Example: Identity**

Identity appears in several concepts:

- Authentication
- Identity Governance
- Federated Identity Management
- Digital Signatures
- Non-Repudiation
- Privileged Access Management

The repeated occurrence of Identity across multiple concepts suggests that it captures a fundamental aspect of cybersecurity knowledge.

**Example  : Traceability**

Traceability appears in:

- Auditing
- Monitoring
- Logging
- Forensics
- Non-Repudiation
- Incident Investigation

Again, the concern is reusable across multiple domains and use cases.

The more broadly a concern is reused across cybersecurity concepts, the more likely it is to represent a foundational semantic element rather than a contextual concept.

### 3.3.4 Atomicity Evaluation

Based on the preceding principles, a Security Concern may be considered a candidate Atomic Security Concern when:

1. It cannot be consistently decomposed into more fundamental Security Concerns.
2. Any attempted decomposition results in loss of security semantics.
3. The concern is reusable across multiple cybersecurity concepts and contexts.

Conversely, a Security Concern should be regarded as composite when it can be represented as a meaningful combination of more fundamental concerns while preserving its cybersecurity semantics.

The following examples illustrate the application of these principles.

**Example: Authentication**

*Authentication= Identity + Authenticity*

Composite Security Concern because it satisfies Principle 1.

**Example: Confidentiality**

*Confidentiality = Secret + Access*

Composite Security Concern because a valid decomposition exists.

**Example: Identity**

No decomposition has been identified that preserves the complete cybersecurity meaning of Identity.

Furthermore, Identity appears in numerous cybersecurity concepts.

So, ***Identity*** can be a Candidate Atomic Security Concern subject to further validation.

**Example: Traceability**

No accepted decomposition has been identified while maintaining the meaning of Traceability.

Traceability is also reused across auditing, monitoring, forensics, investigation, and accountability-related concepts.

So, **Traceability** can be a Candidate Atomic Security Concern subject to further validation.

These principles do not guarantee that a particular concern is atomic. Rather, they provide a systematic basis for evaluating candidate Atomic Security Concerns and distinguishing them from higher-level composite concerns. In this work, atomicity is therefore treated as a research hypothesis rather than an established property. Formal validation methods, composition rules, and mathematical foundations remain subjects for future research.

The next section applies these principles to develop an initial taxonomy of candidate Atomic Security Concerns.

# 4 Taxonomy of Atomic Security Concerns

Having introduced the notions of Security Concern and Atomic Security Concern, the next step is to explore what candidate Atomic Security Concerns might constitute a foundational cybersecurity vocabulary.

The objective of this section is not to establish a definitive or exhaustive taxonomy of Atomic Security Concerns. Rather, it proposes an initial set of candidate concerns intended to illustrate the Atomic Security Concern hypothesis and demonstrate how a broad range of cybersecurity concepts may be represented through compositions of more fundamental concerns.

The candidate concerns presented in this section were identified through the examination of recurring concepts appearing across cybersecurity properties, controls, risks, requirements, capabilities, functions, and architectural patterns. Particular attention was given to concerns that appear repeatedly across different cybersecurity contexts and that appear difficult to decompose without significant loss of security semantics.

The resulting taxonomy should therefore be viewed as illustrative rather than normative. Future research may refine, expand, merge, or redefine individual concerns based on further theoretical and empirical investigation.

## 4.1 Candidate Atomic Security Concerns

Table 1 presents the candidate Atomic Security Concerns proposed in this work.

| **Candidate Atomic Security Concern** | **Description** |
|---|---|
| Identity | The concern that entities can be uniquely distinguished, represented, and referenced within a cybersecurity context. |
| Authenticity | The concern that a claimed identity, source, origin, or assertion is genuine and can be verified. |
| Access | The concern that interactions with resources are permitted, restricted, or denied according to defined conditions. |
| Privilege | The concern that permissions, authorities, or capabilities are assigned to an entity. |
| Secret | The concern that information remains unknown or undisclosed to unauthorized entities. |
| Integrity | The concern that information, systems, processes, or states remain accurate, complete, and unaltered except through authorized actions. |
| Trust | The concern that sufficient confidence exists regarding the expected behavior of an entity, system, process, assertion, or relationship. |
| Traceability | The concern that actions, events, decisions, and state changes can be followed and associated with their origins over time. |
| Evidence | The concern that information exists to support, validate, or demonstrate a claim, action, event, or conclusion. |
| Availability | The concern that systems, services, functions, or information remain usable when required. |
| Recovery | The concern that systems, services, data, or operations can be restored following disruption, failure, or compromise. |

**Table 1. Candidate Atomic Security Concerns**

The concerns listed above should be interpreted as candidate Atomic Security Concerns rather than validated atomic elements. Their inclusion reflects the hypothesis explored in this paper and serves as a basis for investigating how cybersecurity concepts may be represented through composition.

Several observations emerge from the proposed taxonomy.

First, the candidate concerns are intentionally independent of specific technologies, standards, frameworks, or implementation mechanisms. For example, the concern *Identity* remains applicable regardless of whether identities are managed through directories, certificates, federation protocols, or decentralized identity systems.

Second, the proposed concerns are intended to capture cybersecurity semantics rather than cybersecurity mechanisms. Technologies such as encryption, multifactor authentication, logging systems, certificate authorities, or backup platforms are not considered Atomic Security Concerns because they constitute mechanisms that protect, support, enforce, or operationalize underlying concerns.

Third, the candidate concerns appear repeatedly across multiple cybersecurity domains. For example, Identity appears in authentication, identity governance, privileged access management, digital signatures, and non-repudiation. Similarly, Traceability appears across auditing, monitoring, incident response, and forensic investigations.

Finally, the proposed taxonomy should be viewed as a starting point for discussion rather than a finalized classification scheme. Additional concerns may emerge, some concerns may ultimately prove to be composite, and alternative decompositions may arise as the Atomic Security Concern hypothesis is further investigated.

## 4.2 Illustrative Compositions Across Cybersecurity Concepts

The relevance of the proposed taxonomy lies not only in the candidate concerns themselves but also in their potential ability to represent diverse cybersecurity concepts through composition.

**Security Properties**

The first category considered is security properties. Security properties represent desired security outcomes and are commonly used to define protection objectives. The examples below illustrate how such properties may be expressed through combinations of candidate Atomic Security Concerns.

| Security Property | Illustrative ASC Composition |
|---|---|
| Authentication | Identity + Authenticity |
| Confidentiality | Secret + Access |
| Non-Repudiation | Identity + Traceability + Evidence |
| Authorization | Identity + Privilege + Access |
| Accountability | Identity + Traceability + Evidence |

**Table 2. Illustrative Security Properties**

**Security Risks**

Security risks provide a different perspective. Rather than representing desired outcomes, they describe situations in which one or more concerns may be threatened, compromised, or not adequately preserved. The following examples illustrate how risks can be associated with underlying concerns.

| Security Risk | Threat against Illustrative ASC Composition |
|---|---|
| Account Takeover | Identity + Authenticity |
| Unauthorized Data Disclosure | Secret + Access |
| Privilege Escalation | Privilege + Access |
| Log Tampering | Traceability + Evidence |
| Service Outage | Availability |
| Data Corruption | Integrity |

**Table 3. Illustrative Security Risks**

## Security Controls

Security controls represent mechanisms intended to enforce, protect, or strengthen security concerns. The examples below illustrate how controls can be viewed as operational implementations addressing one or more candidate concerns.

| Security Control | Illustrative ASC Composition Enforcement |
|---|---|
| Multi-Factor Authentication (MFA) | Identity + Authenticity |
| Role-Based Access Control (RBAC) | Privilege + Access |
| Digital Signatures | Identity + Authenticity + Evidence |
| Security Logging | Traceability + Evidence |
| Backup and Recovery | Availability + Recovery |
| Certificate Validation | Trust + Authenticity |

**Table 4. Illustrative Security Controls**

## Security Measures and Technologies

Beyond formal controls, cybersecurity relies on technologies and implementation measures that operationalize security objectives. These examples illustrate how concrete technical mechanisms can also be interpreted through compositions of underlying concerns.

| Security Measure | Illustrative ASC Composition Enforcement |
|---|---|
| Encryption | Secret |
| Key Management | Secret + Trust |
| Audit Trails | Traceability + Evidence |
| Disaster Recovery Site | Availability + Recovery |
| Certificate Authority (CA) | Trust + Authenticity |
| Integrity Monitoring | Integrity + Traceability |

**Table 5. Illustrative Security Measures and Technologies**

## Security Functions and Capabilities

Many cybersecurity functions and capabilities encompass broader operational activities than individual controls or technologies. Nevertheless, they can still be interpreted as compositions of candidate concerns. The following examples illustrate this perspective.

| **Security Function / Capability** | **Illustrative ASC Composition** |
|---|---|
| Identity and Access Management (IAM) | Identity + Authenticity + Privilege + Access |
| Privileged Access Management (PAM) | Identity + Privilege + Access + Traceability |
| Security Monitoring | Traceability + Evidence + Integrity |
| Incident Response | Traceability + Evidence + Recovery |
| Key Management | Secret + Trust + Integrity |
| Business Continuity Management | Availability + Recovery |

**Table 6. Illustrative Security Functions and Capabilities**

**Organizational Teams and Specialized Capabilities**

The abstraction is not limited to technical concepts. Organizational teams and specialized capabilities can also be interpreted through the concerns they are responsible for managing, operating, or preserving. The examples below illustrate this possibility.

| **Team / Organizational Capability** | **Illustrative ASC Composition** |
|---|---|
| Identity and Access Management Team | Identity + Authenticity + Privilege + Access |
| Security Operations Center (SOC) | Traceability + Evidence + Integrity |
| Incident Response Team | Traceability + Evidence + Recovery |
| Public Key Infrastructure (PKI) Team | Authenticity + Trust + Secret |
| Disaster Recovery Team | Availability + Recovery |
| Governance, Risk and Compliance (GRC) Capability | Traceability + Evidence + Accountability |

**Table 7. Illustrative Organizational Capabilities**

Several observations emerge from the preceding examples.

First, the same candidate Atomic Security Concerns appear across multiple categories of cybersecurity concepts. For example, *Identity* and *Authenticity* are present in security properties such as Authentication, security risks such as Account Takeover, security controls such as Multi-Factor Authentication, and organizational capabilities such as Identity and Access Management.

Second, security controls, technologies, implementation measures, and organizational capabilities do not necessarily introduce additional foundational concerns. Instead, they appear to protect, support, operationalize, or enforce one or more underlying concerns.

Third, security risks can be interpreted as threats directed toward specific concerns, while security controls and technical measures can be interpreted as mechanisms intended to preserve those same concerns.

These observations support the central hypothesis of this work: a relatively small set of Atomic Security Concerns may provide a common vocabulary capable of representing a broad spectrum of cybersecurity concepts, including properties, risks, controls, technologies, functions, and organizational capabilities.

# 5 Atomic Security Concern as a Foundation of a Cybersecurity Language

The Atomic Security Concern hypothesis extends beyond the identification of candidate foundational concerns. Its broader implication is the possibility that cybersecurity knowledge may be represented through a common language built upon a finite set of foundational concerns.

The objective of this section is not to define a formal cybersecurity language. Rather, it explores how Atomic Security Concerns may provide the vocabulary of such a language and how cybersecurity concepts may emerge through combinations of those concerns.

## 5.1 Atomic Security Concerns as a Cybersecurity Vocabulary

A language requires a vocabulary from which more complex expressions can be constructed. Natural languages rely on words. Chemistry relies on atoms and molecules. Similarly, the Atomic Security Concern hypothesis proposes that cybersecurity knowledge may be expressed through a finite vocabulary of foundational concerns.

Under this perspective, candidate Atomic Security Concerns such as Identity, Authenticity, Access, Privilege, Secret, Integrity, Trust, Traceability, Evidence, Availability, and Recovery do not represent complete cybersecurity concepts. Instead, they provide the elementary cybersecurity vocabulary from which more complex concepts may be expressed.

For example, concepts such as Authentication, Confidentiality, Non-Repudiation, Privileged Access Management, Multi-Factor Authentication, and Digital Signatures can be viewed as different expressions composed from a common vocabulary of concerns.

The significance of this perspective is that cybersecurity knowledge may be described through a relatively small set of recurring concerns rather than through an ever-growing collection of independent concepts.

## 5.2 From Vocabulary to Expressions

If Atomic Security Concerns constitute the vocabulary of a cybersecurity language, then higher-level cybersecurity concepts may be viewed as expressions constructed from that vocabulary.

For example:

*Authentication= Identity + Authenticity*

And,

*Confidentiality = Secret + Access*

And,

*Non-Repudiation = Identity + Traceability + Evidence*

Under this view, cybersecurity concepts are not treated as isolated entities. Instead, they represent different combinations of recurring concerns.

This perspective provides a possible explanation for the observation that many cybersecurity concepts share common foundations despite belonging to different domains, standards, technologies, or organizational practices.

The examples presented in Section 4 suggest that the same vocabulary may be used to represent security properties, security risks, security controls, security measures, organizational capabilities, and specialized security teams. Consequently, concepts that are traditionally described through separate models may be interpreted as different expressions generated from the same underlying vocabulary.

## 5.3 Toward Security Reasoning

One potential implication of representing cybersecurity concepts through Atomic Security Concerns is the ability to reason about relationships among concepts using their constituent concerns.

For example, two cybersecurity artifacts may share one or more concerns. Such common concerns may indicate conceptual overlap, shared objectives, or complementary relationships between the artifacts.

Consider the following examples:

*Authentication = Identity + Authenticity*

And,

*Multi-Factor Authentication = Identity + Authenticity*

Both concepts share the concerns Identity and Authenticity. This observation explains why Multi-Factor Authentication contributes directly to strengthening Authentication.

Similarly:

*Unauthorized Data Disclosure = Threat against Secret + Access*

And,

*Encryption= Protection of Secret*

Both concepts are related through the concern Secret, suggesting a direct conceptual relationship between the risk and the measure intended to mitigate it.

More generally, concepts represented through Atomic Security Concerns may be compared according to:

### 5.3.1 Shared Concerns

Two concepts may share common concerns.

Illustratively:

*Authentication = Identity + Authenticity*

And,

*Identity Governance = Identity + Privilege + Traceability*

Both concepts share the concern:

Identity

which suggests a semantic relationship between them.

### 5.3.2 Aggregated Concerns

Multiple concepts may collectively contribute to a broader set of concerns.

For example:

*Encryption = Secret*

And,

*Access Control = Access*

Together they contribute to:

*Secret + Access*

which corresponds to the concerns represented by Confidentiality.

### 5.3.3 Inclusion Relationship

Some concepts may encompass all concerns represented by another concept.

For example:

*Authentication = Identity + Authenticity*

And,

*Identity and Access Management = Identity + Authenticity + Privilege + Access*

The concerns associated with Authentication are entirely contained within those associated with Identity and Access Management, suggesting that Authentication forms part of the broader IAM capability.

These examples illustrate how cybersecurity concepts may potentially be analyzed through their underlying concerns rather than exclusively through framework-specific terminology. Although this work does not introduce a formal reasoning model, the proposed perspective suggests that concern-based representations may support more explicit and traceable relationships among cybersecurity concepts.

## 6 Future Work

This work introduces the Atomic Security Concern hypothesis as a conceptual foundation for investigating whether cybersecurity knowledge can be expressed through a finite set of foundational concerns. Several directions for future research emerge from this initial exploration.

A first direction consists of refining and validating the proposed taxonomy of candidate Atomic Security Concerns. This includes evaluating the completeness of the taxonomy, investigating alternative decompositions, and assessing whether additional concerns are required to represent cybersecurity concepts across different domains, architectures, and frameworks.

A second direction involves the formalization of the proposed concepts and their relationships. Future work will investigate formal definitions of Atomic Security Concerns, composition mechanisms, and concern-based reasoning using mathematical foundations. Such formalization may enable the development of a rigorous conceptual model in which cybersecurity concepts can be represented, analyzed, and compared through well-defined operations.

A third direction consists of exploring the construction of a formal cybersecurity language founded on Atomic Security Concerns. In such a language, Atomic Security Concerns would constitute the vocabulary, composition rules would define the syntax, and the meaning associated with concerns and their compositions would provide the semantics. This may ultimately support formal reasoning, knowledge representation, ontology development, automated analysis, and AI-assisted cybersecurity decision-making.

## 7 Conclusion

This paper introduced the concepts of Security Concerns and Atomic Security Concerns (ASCs) and explored the hypothesis that cybersecurity knowledge may be represented through a finite set of foundational concerns. An initial taxonomy of candidate Atomic Security Concerns was proposed, together with illustrative examples demonstrating how cybersecurity properties, risks, controls, measures, capabilities, and organizational functions may be represented as compositions of recurring concerns.

The proposed perspective suggests that Atomic Security Concerns may provide a common vocabulary capable of describing cybersecurity concepts that are traditionally represented through different abstraction models and frameworks. By focusing on underlying concerns rather than framework-specific constructs, the approach offers a potential foundation for a more consistent, traceable, and interoperable representation of cybersecurity knowledge.

The contribution of this work is primarily conceptual. Rather than formally proving the existence or atomicity of individual concerns, it introduces a research hypothesis and an initial framework for investigating the foundations of a cybersecurity language. If further validated and formalized, Atomic Security Concerns may ultimately serve as the vocabulary of a formal cybersecurity language, with composition mechanisms providing its syntax and concern semantics providing its meaning. Such a language could offer a unified basis for representing, reasoning about, and communicating cybersecurity knowledge across domains, frameworks, and organizational contexts.